\documentclass[a4paper]{article}

\usepackage{INTERSPEECH2020}
\usepackage[utf8]{inputenc}

\usepackage[unicode]{hyperref}
\hypersetup{bookmarksopen=true,bookmarksnumbered=true}

\hypersetup{
	colorlinks=true,
	linkcolor=black,
	citecolor=black,
	filecolor=black,
	urlcolor=black
}

\title{Speaker dependent articulatory-to-acoustic mapping \\ using real-time MRI of the vocal tract}
\name{Tamás Gábor Csapó$^{1,2}$}
\address{
  $^1$Department of Telecommunications and Media Informatics, \\
	Budapest University of Technology and Economics, Budapest, Hungary \\
	$^2$MTA-ELTE Lendület Lingual Articulation Research Group, Budapest, Hungary}
\email{csapot@tmit.bme.hu}

\begin{document}

\maketitle
\begin{abstract}

Articulatory-to-acoustic (forward) mapping is a technique to predict speech using various articulatory acquisition techniques (e.g. ultrasound tongue imaging, lip video). Real-time MRI (rtMRI) of the vocal tract has not been used before for this purpose. The advantage of MRI is that it has a high `relative' spatial resolution: it can capture not only lingual, labial and jaw motion, but also the velum and the pharyngeal region, which is typically not possible with other techniques. In the current paper, we train various DNNs (fully connected, convolutional and recurrent neural networks) for articulatory-to-speech conversion, using rtMRI as input, in a speaker-specific way. We use two male and two female speakers of the USC-TIMIT articulatory database, each of them uttering 460 sentences. We evaluate the results with objective (Normalized MSE and MCD) and subjective measures (perceptual test) and show that CNN-LSTM networks are preferred which take multiple images as input, and achieve MCD scores between 2.8--4.5~dB. In the experiments, we find that the predictions of speaker `m1' are significantly weaker than other speakers. We show that this is caused by the fact that 74\% of the recordings of speaker `m1' are out of sync.

\end{abstract}

\section{Introduction}

The articulatory movements are directly linked with the acoustic signal in the speech production process. A long-standing issue in articulatory research concerns the estimation of vocal tract configuration and/or its mapping with the acoustic parameters. There are several available methods to model the relation of articulatory movements and the resulting speech signal, mostly referred as `articulatory-to-acoustic (forward) mapping'. 
One of the target applications is often called as `Silent Speech Interface' (SSI~\cite{Denby2010}). This has the main idea of recording the soundless articulatory movement, and automatically generating speech from the movement information, while the subject is not producing any sound.
For the automatic conversion task, typically electromagnetic articulography (EMA)~\cite{Wang2012a,Cao2018,Taguchi2018,Illa2019}, ultrasound tongue imaging (UTI)~\cite{Denby2004,Hueber2011,Jaumard-Hakoun2016,Csapo2017c,Grosz2018,Toth2018,Moliner2019,Csapo2019}, permanent magnetic articulography (PMA)~\cite{Fagan2008,Gonzalez2017a}, surface electromyography (sEMG)~\cite{Janke2017,Wand2018}, Non-Audible Murmur (NAM)~\cite{Shah2018} or video of the lip movements~\cite{Akbari2018} are used. According to our knowledge, magnetic resonance imaging (MRI), which provides detailed information about the vocal tract, has not been used yet for articulatory-to-acoustic mapping.

Recently, significant advances in MR research (software, hardware, and reconstruction strategies) have allowed real-time MRI (rtMRI) to be a powerful modality for speech production research and for investigating the movement of the articulators~\cite{Narayanan2014,Ramanarayanan2018,Toutios2019}. The advantage of rtMRI is that it provides dynamic information about the full midsagittal plane of the upper airway, even during continuous spoken utterances. It can capture not only lingual, labial and jaw motion, but also the articulation of the velum and the pharyngeal region, which is typically not possible with other articulatory acquisition techniques. Besides, such imaging data helps to comprehend the generation of coronal, pharyngeal, and nasal segments. The sampling rates of rtMRI are relatively low (around 20~fps), but are acceptable for running speech. A disadvantage is the large background noise in speech recordings, but noise cancellation can yield an acceptable speech signal, which can be synchronized to the articulatory signal. Also, the presence of a substantial number of artifacts and noise make automatic extraction and interpretation of features a difficult problem. Overall, rtMRI provides high relative spatial information in the midsagittal view with relatively low temporal resolution~\cite{Toutios2019}; therefore it is a potentially suitable technique for articulatory-to-acoustic conversion.

Several studies have applied MRI for articulatory-related speech technologies: e.g.\ articulatory speech synthesis~\cite{Alexander2019,Toutios2016}, articulatory video synthesis~\cite{SChandana2018}, speech recognition using acoustic and articulatory data~\cite{Douros2018}, acoustic-to-articulatory inversion~\cite{Li2015}, phoneme classification from articulation~\cite{Katsamanis2011,Saha2018,Douros2018,VanLeeuwen2019}, while there has been no research yet on MRI-based direct articulatory-to-speech synthesis. Some of the above studies are more relevant for our scenario, and we discuss them in detail.
Li et al.\ present a system for acoustic-to-articulatory inversion using midsagittal rtMRI, where restricted Boltzmann machine, GMM and linear regression are applied for the mapping~\cite{Li2015}. In this inversion task, the input of the machine learning models are acoustic feature vectors (24-order line spectral pairs, with a context window of 10 acoustic frames), whereas the target is the gray value vectors of the 68$\times$68~pixel MR images. According to the results, deep architectures are able to obtain better inversion accuracy than the GMM-based method, in terms of RMSE~\cite{Li2015}. However, only a single speaker (`f1') was used from the USC-TIMIT database~\cite{Narayanan2014}.
Katsamanis et al.\ conducted the first large-scale articulatory recognition experiment using extracted vocal tract shapes outlined from MR images~\cite{Katsamanis2011}. HMM-based recognition was applied and resulted in nearly 50\% accuracy when using 30 classes, with the data of one male speaker from the USC-TIMIT database~\cite{Narayanan2014}. 
Saha and his colleagues experimented with identifying different vowel-consonant-vowel (VCV) sequences from dynamic shapings of the vocal tract recorded using MRI, as a step towards subject-invariant automatic mapping of vocal tract shape geometry to acoustics~\cite{Saha2018}. They used Long-term Recurrent Convolutional Networks (including a pre-trained ResNet50) models, which makes the network spatiotemporally deep enough to capture the sequential nature of the articulatory data. The context of 16 MRI frames is used as input, and vowels / consonants / VCV sequences are the target. The rtMRI data of 17 speakers (9 female and 8 male) was used from the USC Speech and Vocal Tract Morphology MRI Database~\cite{Sorensen2017}. The final classification accuracies are relatively low (42\% for VCV), which could be explained by the fact that similar articulatory movements are getting mapped to different sounds, which is a vital issue that the LRCN algorithm is unable to detect~\cite{Saha2018}.
Van Leeuwen and his colleagues trained a convolutional neural network (CNN) for the classification of 27 different sustained phonemes~\cite{VanLeeuwen2019}, using 17 speakers of the USC Speech and Vocal Tract Morphology MRI Database~\cite{Sorensen2017}. Although the top-1 accuracy is only 57\%, this can be explained by the small dataset used for training: only a single midsagittal slice of the 3D MRI recordings was used for each speaker and phoneme, resulting in 489 images. Besides the classification experiments, they show saliency maps, which provide new insights on what the CNN `sees', and reveal that the network has learned to focus on those parts of the images that represent the crucial articulatory positions needed to distinguish the different phonemes~\cite{VanLeeuwen2019}.


Based on this overview, rtMRI data has not been used previously for direct articulatory-to-acoustic mapping. In the current paper, we train various neural network architectures (fully connected, convolutional, and recurrent neural networks) for articulatory-to-speech conversion, using real-time magnetic resonance images of the vocal tract in a speaker-specific way.

\section{Methods}

\subsection{Data}

\begin{figure}
\centering
\includegraphics[trim=0.2cm 0.2cm 0.2cm 0.2cm, clip=true, width=\columnwidth]{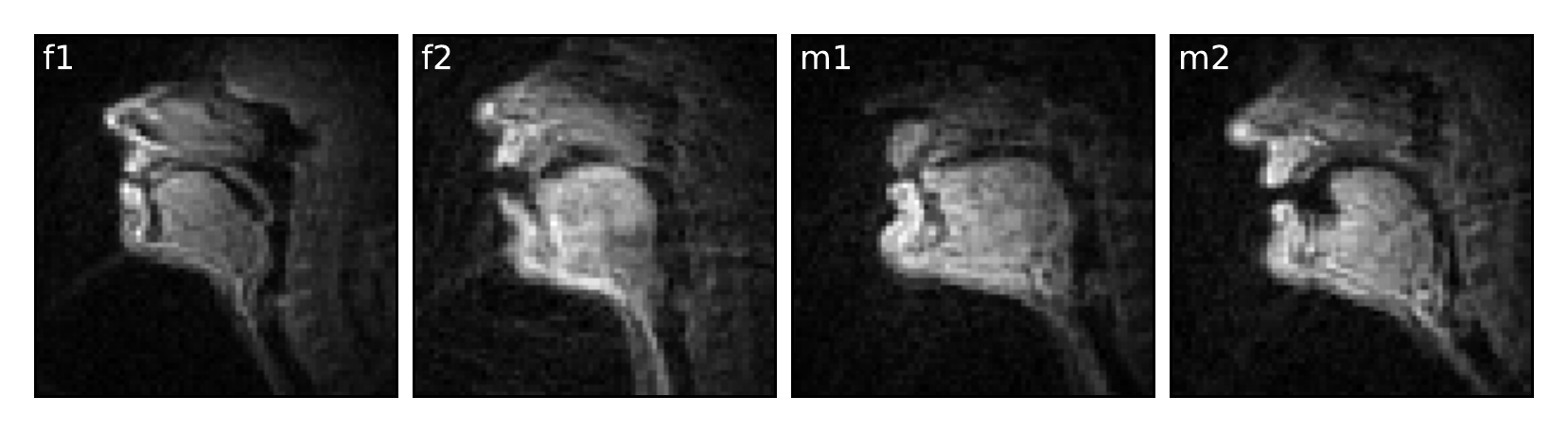}
\vspace{-4mm}
\caption{Sample MRI data from the four speakers.}
\label{fig:mri_sample}
\vspace{-4mm}
\end{figure}

We used two male (`m1' and `m2') and two female (`f1' and `f2') speakers from the freely available USC-TIMIT MRI database \cite{Narayanan2014}. This contains large-scale data of synchronized audio and rtMRI for speech research, from American English subjects. The vocal tracts were imaged in the mid-sagittal plane while lying supine and reading 460 MOCHA-TIMIT sentences. The MRI data were acquired using a Signa Excite HD 1.5T scanner with an image resolution in the mid-sagittal plane of 68$\times$68~pixels (2.9$\times$2.9mm). Fig.~\ref{fig:mri_sample} shows sample MRI data. The image data were reconstructed at 23.18~frames/second. The audio was simultaneously recorded at a sampling frequency of 20~kHz inside the MRI scanner while subjects were imaged. Noise cancellation was also performed on the acoustic data. 

\subsection{Vocoder}

To create the speech synthesis targets, we encoded the audio recordings using an MGLSA vocoder~\cite{Imai1983} at a frame shift of 1 / (23.18 fps) = 863 samples and with 1024 samples frame size, which resulted in F0 and 24-order spectral (MGC-LSP) features. The spectral parameters served as the training targets of the DNN, while we used the original (inverse filtered) LP residual excitation for the final synthesis.

\subsection{Deep neural network architectures}

In our earlier studies on ultrasound-to-speech synthesis, we were using fully-connected feed-forward neural networks (FC-DNN)~\cite{Csapo2017c,Grosz2018,Toth2018}, CNNs~\cite{Moliner2019,Csapo2019} and recurrent neural networks~\cite{Moliner2019}. Here, we test similar network types, without automatic hyperparameter optimization. In all cases and for all speakers, we trained speaker-specific models, and we split the data into 430 sentences for training, 20 sentences for validation, and 10 sentences for testing. We used Adam optimizer, trained the networks for 100 epochs, and applied early stopping with a patience of 5 on the validation loss. The input MRI pixels were scaled to [0-1], while the target spectral features were normalized to zero mean, unit variance. The data is passed to the networks in batches of 128 frames. The cost function applied for the MGC-LSP regression task was the MSE.

\subsubsection{FC-DNN (baseline)}

In the simplest case, we trained FC-DNNs with 5 hidden layers, each hidden layer consisting of 1000 neurons, with ReLU activation. The input layer consisted of 4\,624 neurons (taking the raw pixel values of MRI). The output layer was a linear one, with one neuron for each MGC-LSP feature.

\subsubsection{CNN}
\label{sec:CNN}

Next, we tested CNNs, as typically, they are more suitable for the processing of images than simple FC-DNNs. All CNNs had one 68$\times$68 pixel MR image as input and had the same structure: three convolutional layers (kernel size: 3$\times$3, number of filters: 8, 16 and 32, respectively), each followed by max-pooling. Finally, two dense layers were used with 500 neurons each. In all hidden layers, ReLU activation was used.

\subsubsection{CNN-LSTM}

\begin{figure}
\centering
\includegraphics[trim=0.0cm 18.5cm 11.0cm 0.0cm, clip=true, width=0.8\columnwidth]{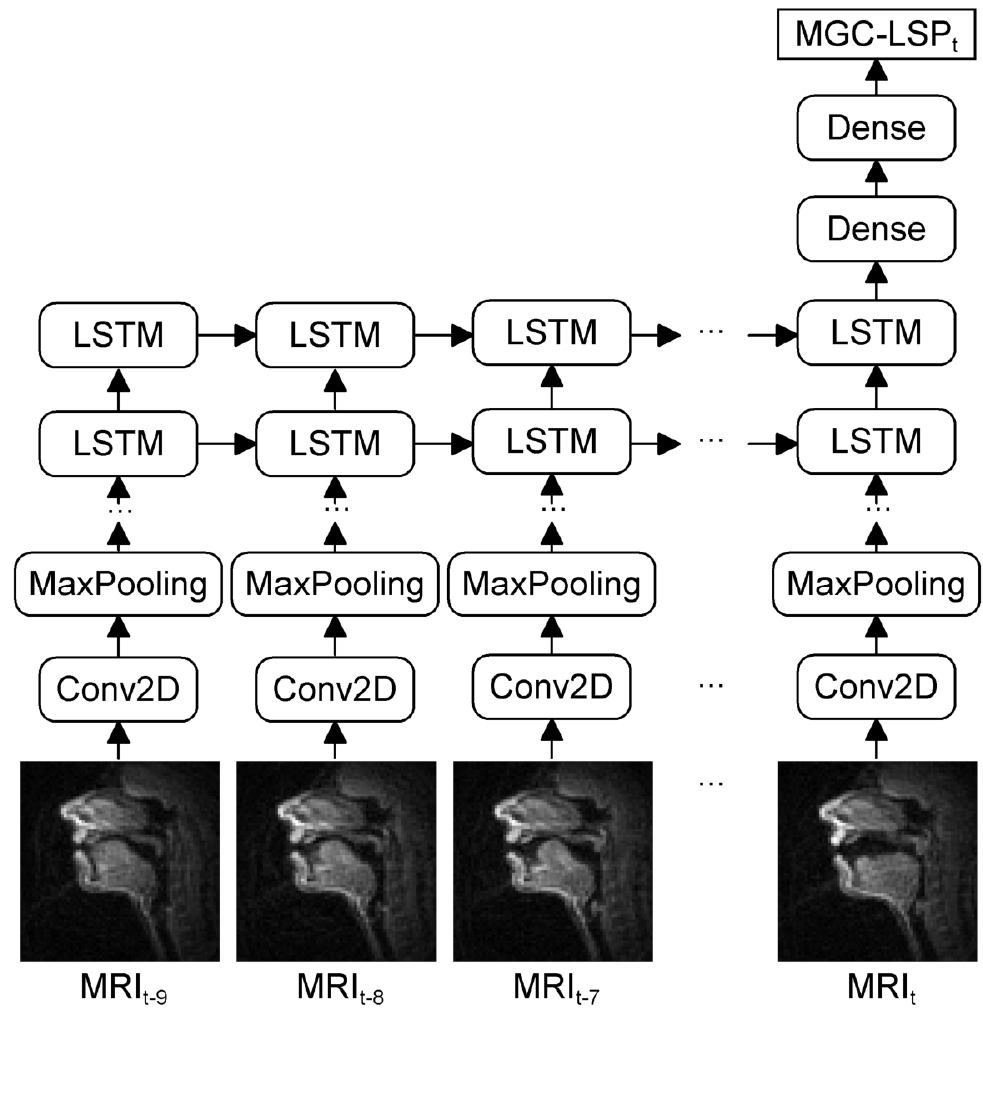}
\vspace{-4mm}
\caption{Block diagram of the CNN-LSTM network.}
\label{fig:cnn_lstm}
\vspace{-4mm}
\end{figure}

Also, we hypothesized that using multiple consecutive images as input can increase the accuracy of the regression. 
The most ambitious network in this work is a recurrent one consisting of a combination of CNNs and Long Short-Term Memory units (LSTMs). The motivation for designing this network comes from the fact that, in \cite{Csapo2017c,Grosz2018,Moliner2019}, we achieved better results when using consecutive ultrasound frames. The presented network is aimed to explore this correlation between consecutive MRI frames.
Figure \ref{fig:cnn_lstm} shows the architecture of the CNN-LSTM network, consisting of three distinguished parts: three CNN layers, two LSTM layers, and a fully connected end. The CNN part is the same as in Sec.~\ref{sec:CNN}. It is followed by two LSTM layers, having 500 neurons, and ReLU activation for each. We use a sequence size of 10 (accounting for roughly 430~ms of the input articulatory data) in order to incorporate time information. In the end, a fully connected network is added, similarly to the CNN approach above. To keep the FC-DNN (baseline) and the CNN-LSTM comparable with respect to parameter count, both models have approximately 8.6 million tunable parameters.


\section{Results}

After training the neural networks, the prediction accuracy was evaluated on the test set (10 sentences for each speaker). We synthesized sentences by filtering the original LP residual excitation using an MGLSA vocoder applying the DNN-predicted MGC-LSP features.

%
%
%

\subsection{Objective evaluation}
\label{sec:objective}

\begin{table}
\caption{NMSE scores on the validation and test set.} \label{tab:objective_NMSE}
\vspace{-2mm}
\centering
\begin{tabular}{l||c|c|c}
        & \multicolumn{3}{c}{{Normalized MSE (validation / test)}} \\
\cline{2-4}
speaker & FC-DNN & CNN & CNN-LSTM \\
\hline\hline
`f1' & 0.49 / 0.51 & 0.47 / 0.48 & 0.32 / 0.33 \\
`f2' & 0.49 / 0.53 & 0.48 / 0.47 & 0.31 / 0.35 \\
`m1' & 0.72 / 0.80 & 0.70 / 0.88 & 0.67 / 0.87 \\
`m2' & 0.45 / 0.46 & 0.44 / 0.46 & 0.31 / 0.34 \\
\hline
average & 0.54 / 0.58 & 0.52 / 0.57 & 0.40 / 0.47 \\

\end{tabular}

\vspace{2mm}
\caption{MCD scores on the test set.} \label{tab:objective_MCD}
\vspace{-2mm}
\centering
\begin{tabular}{l||c|c|c}
        & \multicolumn{3}{c}{{Mel-Cepstral Distortion (dB)}} \\
\cline{2-4}
speaker & ~FC-DNN~ & ~~~~CNN~~~~ & CNN-LSTM \\
\hline\hline
`f1' & 3.63 & 3.49 & 2.82 \\
`f2' & 5.33 & 5.14 & 4.40 \\
`m1' & 4.32 & 4.49 & 4.34 \\
`m2' & 5.41 & 5.33 & 4.51 \\
\hline
average & 4.67 & 4.61 & 4.02 \\

\end{tabular}

\end{table} 

On the validation set and on the synthesized sentences (being the test set), we first measured the Mean Square Error (MSE) between the original and predicted MGC-LSP features. The calculations were done on the normalized (zero mean, unit variance) features, as the MGC-LSP target values varied at different scales, otherwise, the output having the largest range (MGC-LSP (0)) would have dominated the MSE error. The normalized MSE values calculated on the validation and test are shown in Table~\ref{tab:objective_NMSE}, separately for each speaker. Overall, the tendencies are the same for all speakers: the weakest network seems to be the baseline FC-DNN (test NMSE: 0.58), followed by the CNN (test NMSE: 0.57), and finally, the CNN-LSTM having the smallest error (test NMSE: 0.47). The difference between the FC-DNN and CNN is negligible, whereas the CNN-LSTM is significantly better than these two systems. Interestingly, speakers `f1'/`f2'/`m2' have scores in a similar range, while the results for speaker `m1' are much weaker (roughly twice as bad).

The other metric chosen in this test is the Mel-Cepstral Distortion (MCD), following Kubichek~\cite{Kubichek1993}. This metric is a standard way to evaluate text-to-speech synthesis systems. In general, the advantage of MCD is that it is better correlated with perceptual scores than other objective measures~\cite{Kubichek1993}. Table~\ref{tab:objective_MCD} shows the MCD results in dB for all the neural networks and speakers. The lowest MCD result, with an average value of 4.02~dB, belongs to the CNN-LSTM network, which is the most complex network of our study. Similarly to the NMSE, the FC-DNN and CNN networks resulted in higher errors (in terms of MCD), but in case of the Mel-Cepstral Distortion, the advantage of CNNs over FC-DNNs is also visible (average MCD: 4.61~dB vs.~4.67~dB). The MCD measure does not show as strong dependency on the speakers as it was the case with the NRMS, because the MCD calculation does not include the 0th coefficient of MGC-LSP. The smallest MCD results were achieved with speaker `f1', with all three networks. Speaker `m2' achieved the highest MCD scores (indicating low spectral similarity), which does not correlate with the same speaker's MSE values.

According to these objective experiments, both measures have shown the advantage of using recurrent networks (namely, CNN-LSTM), instead of the networks which are taking single images as input (the FC-DNN and CNN types).

\subsection{Subjective listening test}

In order to determine which proposed system is closer to natural speech, we conducted an online MUSHRA (MUlti-Stimulus test with
Hidden Reference and Anchor) listening test~\cite{mushra}. The advantage of MUSHRA is that it allows the evaluation of multiple samples in a single trial without breaking the task into many pairwise comparisons. Our aim was to compare the natural sentences and a vocoded reference with the synthesized sentences of the baseline, the proposed approaches and a lower anchor (the latter having white noise excitation during resynthesis). We included cases where the original LP residual was used as excitation signal and the DNN-predicted MGC-LSP spectral features were used during synthesis. In the test, the listeners had to rate the naturalness of each stimulus in a randomized order relative to the reference (which was the natural sentence), from 0 (very unnatural) to 100 (very natural). We chose four sentences from the test set of each speaker (altogether 16 sentences). The samples can be found at \url{http://smartlab.tmit.bme.hu/interspeech2020_mri2speech}.

\begin{figure}
\centering
\includegraphics[trim=0.25cm 0.3cm 0.2cm 0.25cm, clip=true, width=\columnwidth]{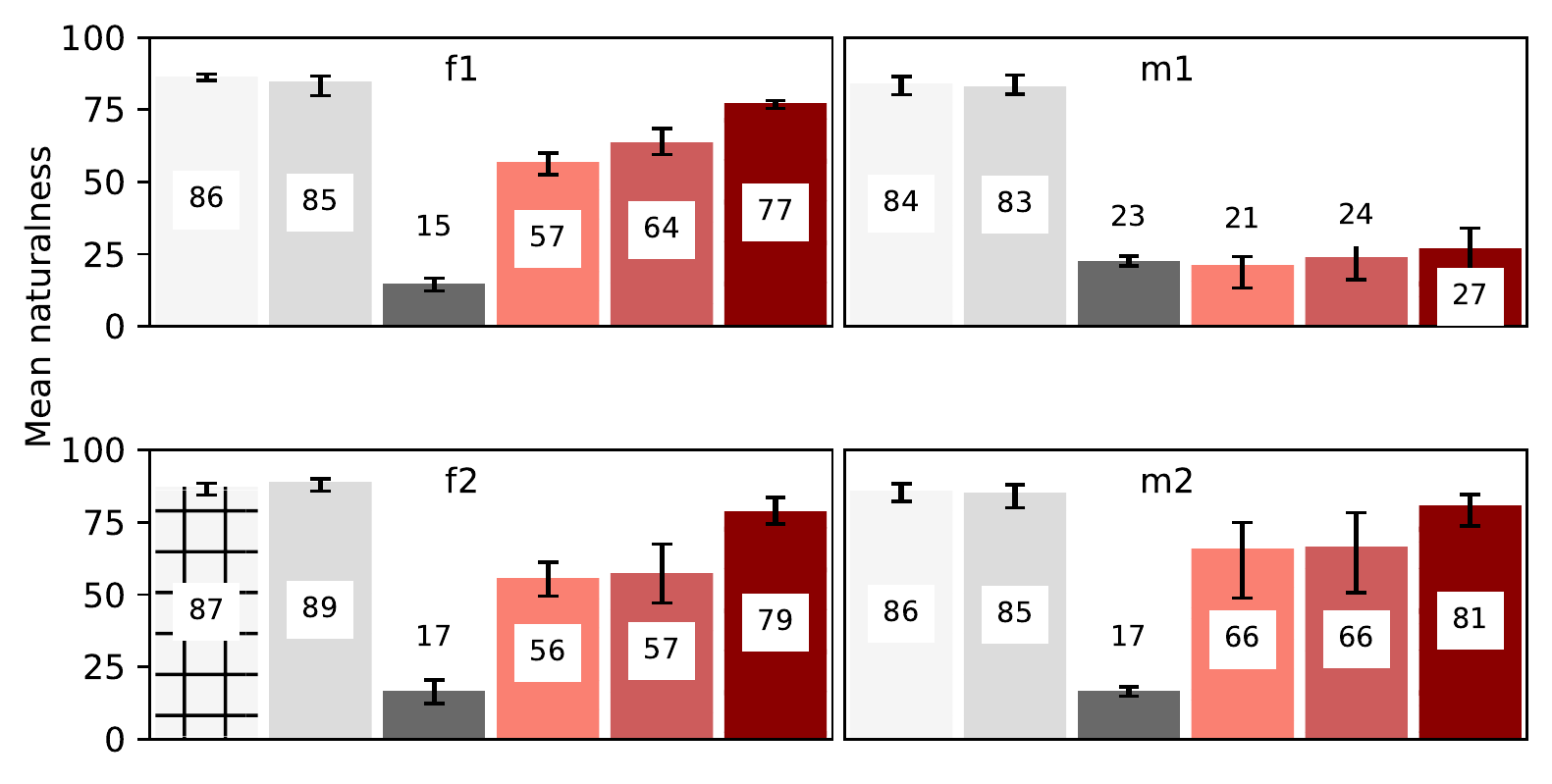}
\includegraphics[trim=0.25cm 0.4cm 0.2cm 0.25cm, clip=true, width=\columnwidth]{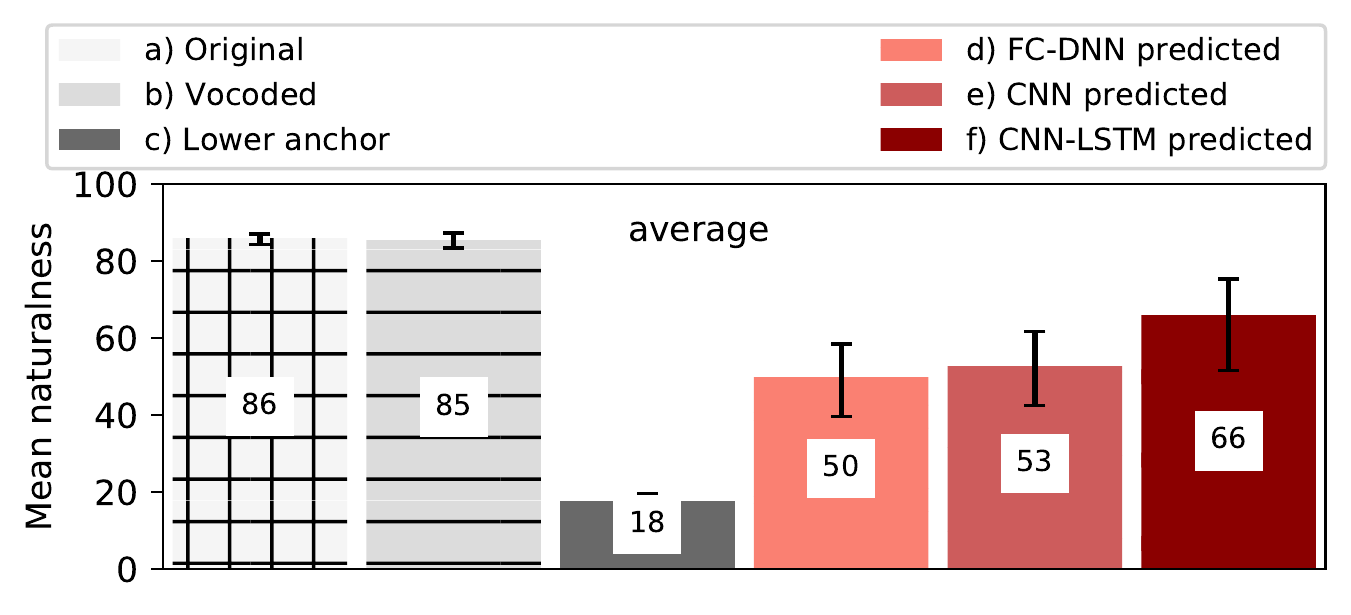}
\caption{\textit{Results of the subjective evaluation for the naturalness question, speaker by speaker (top) and average (bottom). The errorbars show the 95\% confidence intervals.}}
\label{fig:results_mushra}
\end{figure}

Each sentence was rated by 10 subjects (one native English; 4~females, 6 males; 22--44 years old; not including the author). On average, the test took 16 minutes to complete. Fig.~\ref{fig:results_mushra} shows the average naturalness scores for the tested approaches. The lower anchor version (with white noise excitation) achieved the weakest scores, while the original sentences and vocoded re-synthesis were rated the highest, as expected. As the original recordings contained significant background noise and echoes, in several cases, the vocoded version was rated higher than the original. In general, FC-DNN and CNN were rated as roughly equal, while the CNN-LSTM was preferred.
To check the statistical significance of the differences, we conducted Mann-Whitney-Wilcoxon rank-sum tests with a 95\% confidence level, showing that the CNN-LSTM was significantly preferred both over FC-DNN and CNN. The CNN ranked slightly higher than the FC-DNN, but this difference is not statistically significant. When checking the speaker by speaker results, we can see that `m1' clearly ranked lower scores than the other three speakers (similarly to the NMSE measure). In his case, the differences between FC-DDN, CNN, and CNN-LSTM are not significant.

As a summary of the listening test, a clear preference towards the recurrent neural network could be observed, indicating that having multiple images as input is advantageous.

\section{Discussion}

In general, recurrent neural networks are more suitable to process sequential data than convolutional or simple fully-connected networks. As Saha and his colleagues compare for MRI-based phone recognition~\cite{Saha2018}, typical algorithms for video-based action recognition are: 3D Convolutional Networks, Two stream Convolutional Networks, and Long-term Recurrent Convolutional Networks. All of these algorithms incorporate spatial and temporal feature extraction steps to capture complementary information from the individual consecutive still frames as well as between the frames, which is a key for the processing of sequential data. In the current study, we have shown that CNN-LSTMs are more suitable to process MR images than FC-DNNs and (2D) CNNs.



\begin{figure}
\centering
\includegraphics[trim=1.0cm 1.05cm 2.0cm 2.0cm, clip=true, width=\columnwidth]{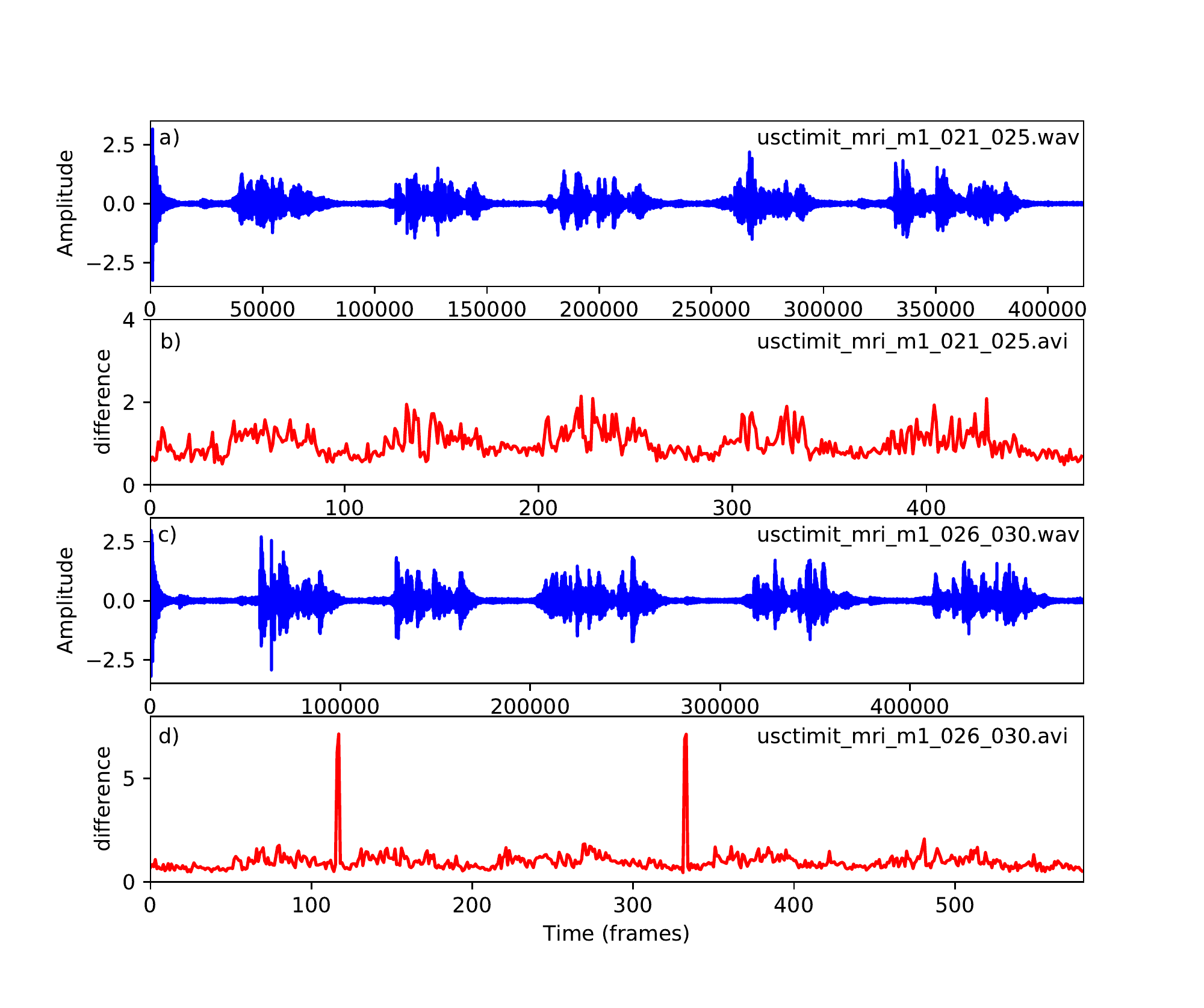}
\caption{Waveform (a and c) and frame-by-frame MRI pixel difference (b and d) of two sentences from speaker `m1'.}
\label{fig:MRI_check_diff}
\end{figure}

The surprisingly high errors and low subjective evaluations of speaker `m1' cannot be explained by the properties of the neural networks, as the circumstances (e.g.\ amount of training data and hyperparameters of the networks) were the same for each speaker. After the above experiments, we checked the audio and visual data of speaker `m1', and found that in many cases, the MRI data and audio are out of sync, sometimes just in the middle of the recordings. To check this automatically, for each sequence of MR images, we calculated the frame-by-frame pixelwise absolute difference. A sample for this is shown in Fig.~\ref{fig:MRI_check_diff} (note that the data was recorded in batches of five sentences). The sharp peaks in the bottom (d) subfigure show potential misalignments of consecutive frames. By listening to this recording, audiovisual misalignment can be found at frame 117 (5.05~s) and 333 (14.37~s). According to these simple measurements, 68 recordings out of 92 (74\% of the data) are out of sync, which explains the weak scores of speaker `m1'.
A future plan is to use more advanced solutions for reconstructing the audiovisual synchrony~\cite{Eshky2019}.

We can compare the results of this study to earlier articulatory-to-acoustic mapping experiments that were using other imaging techniques.
For ultrasound-to-speech conversion, the typical values of MCD were around 5~dB~\cite{Moliner2019}, whereas here we achieved MCD around 2.8--4.5~dB. The reason might be that ultrasound can only capture the movement of the tongue, and higher frame rates (around 100~fps) cannot compensate the lower relative spatial information. Lip-to-speech mapping is a significantly more difficult task, as lip movement shows less information than the full articulation~\cite{Akbari2018}. Usually, the lip-to-speech synthesized sentences are less intelligible than ultrasound.
Although in the USC-TIMIT dataset the resolution of the MR images is only 68$\times$68~pixels, this allows for a larger `relative' spatial resolution than ultrasound or lip images, as MRI can visualize the structure of the whole vocal tract. Our experiments have shown that this is clearly an advantage of rtMRI, and the high `relative' spatial resolution is more important than the relatively low (around 20~fps) time resolution.

\section{Conclusions}

In this work, we used midsagittal rtMRI images of the vocal tract for articulatory-to-acoustic mapping. We applied FC-DNNs, convolutional, and recurrent neural networks, and have shown that CNN-LSTMs are the most suitable for this task.

The input real-time MR images have a relatively low spatial and temporal resolution (but high `relative' spatial resolution), and are infested with noises and reconstruction artifacts~\cite{Saha2018}. In our work, we were using raw MR images and did not apply any preprocessing. However, noise and artifact reduction on the input images might enhance the accuracy of the mapping.

Although rtMRI is not suitable for the potential application of a Silent Speech Interface, as it is not portable, our methods are a kind of scientific exploration, and the articulatory-to-acoustic results shown above might be useful for other modalities having similar properties (e.g.\ ultrasound and lip images).

The keras implementations are accessible at \url{https://github.com/BME-SmartLab/mri2speech}.

\section{Acknowledgements}

The author was funded by the National Research, Development and Innovation Office of Hungary (FK 124584 and PD 127915 grants). The Titan X GPU for the deep learning experiments was donated by the NVIDIA Corporation. We would like to thank USC for providing the USC-TIMIT articulatory database.

\clearpage

\bibliographystyle{IEEEtran}

\bibliography{ref_collection_csapot_nourl}

\end{document}